\documentclass[preprint,preprintnumbers,amsmath,amssymb]{revtex4}
\usepackage{graphicx}
\usepackage{epsfig}
\usepackage{lscape}
\usepackage{dcolumn}
\usepackage{bm}
\begin{document}
\title{ \textit{Ab initio} studies of structures and properties of small potassium clusters}
\author{Arup Banerjee$^{a}$, Tapan K. Ghanty$^{b}$, and  Aparna Chakrabarti$^{c}$\\
(a) Laser Physics Application Division, Raja Ramanna Centre for Advanced Technology\\
Indore 452013, India\\
(b) Theoretical Chemistry Section, Chemistry Group, \\
Bhabha Atomic Research Centre, Mumbai 400 085, India\\
(c) Semiconductor Laser Section, Raja Ramanna Centre for Advanced Technology\\
Indore 452013, India}
\begin{abstract}
We have studied the structure and properties of potassium clusters containing 
even number of atoms ranging from 2 to 20 at the \textit{ab initio} level. The geometry optimization calculations are performed using all-electron density functional theory with gradient corrected exchange-correlation functional. Using these optimized geometries we investigate the evolution of binding energy, ionization potential, and static polarizability with the increasing size of the clusters. The polarizabilities are calculated by employing  M$\ddot{o}$ller-Plesset perturbation theory and time dependent density functional theory. The polarizabilities of dimer and tetramer are also calculated by employing large basis set coupled cluster 
theory with single and double excitations and perturbative triple excitations. The time dependent density functional theory calculations of polarizabilities are carried out with two different exchange-correlation potentials: (i) an asymptotically correct model potential and (ii) within the local density approximation.  A systematic comparison with the other available theoretical and experimental data for various properties of small potassium clusters mentioned above has been performed. These comparisons reveal that both the binding energy and the ionization potential obtained with gradient corrected potential match quite well with the already published data. Similarly, 
the polarizabilities obtained with M$\ddot{o}$ller-Plesset perturbation theory and with model potential are quite close to each other and also close to experimental data. 
\end{abstract}
\maketitle 
\section{Introduction}
During last two decades rapid progress in the experimental methods of producing atomic and molecular clusters in controlled fashion along with development of sophisticated theoretical tools to handle such finite fermionic systems at the \textit{ab initio} level led to emergence of the field of cluster science as one of the most exciting and productive discipline of physics, chemistry, and material science.  Metal clusters, specially those of alkali-metal atoms Li, Na, and K played an important role in the development of cluster physics as a branch of modern physics and chemistry. Interest in the study of alkali-metal clusters  grew with the pioneering work of Knight and co-workers \cite{knight} . These researchers discovered that certain clusters, those with magic number 8, 20, 34, 40, $\cdots$ of atoms are more stable and consequently were found more abundantly in the mass spectra of these clusters. The existence of magic number clusters is attributed to electronic shell structure of the clusters. Other properties like ionization potential, electron affinity, and static polarizabilities of metal clusters also show significance of the shell structure. Besides these, photoabsorption cross sections have also been measured for alkali metal clusters and have been investigated theoretically at various levels. A large body of theoretical work on the electronic structure and optical response properties of alkali-metal clusters exists in the literature. Majority of the theoretical work have been carried out by employing Density functional theory (DFT) and time dependent DFT (TDDFT) within the spherical jellium background model (SJBM) (see the review article \cite{brackreview,alonso}). The SJBM replaces the discrete ionic structure of clusters by a spherically symmetric uniform positive charge background thus making it possible to carry out calculations for the optical response properties of reasonably large clusters of around 100 atoms \cite{brackreview,madjet}. In last ten years or so, several all electron \textit{ab initio}  calculations devoted to the ground state and the optical response properties of sodium clusters taking into account the actual geometrical arrangement of the sodium atoms  have been reported in the literature \cite{moulett,andreoni,guan,calmanici,kummel,pacheco3,solovyov,kronik,ghanty1,jiemchooroj,arup1}. 
However, these calculations could handle clusters with sizes smaller than those that could be studied by performing jellium based calculations. We should mention here that the calculations of structure, electronic and static polraizability of small sodium and lithium clusters have also been performed
by employing \textit{ab initio} correlated wavefunction based methods like MP2, MP4, and SCF-CI\cite{ghanty1,ghanty2,koutockey}. 

We note here that among these numerous studies involving properties of alkali-metal clusters only very few are devoted to the calculations of properties of potassium atom clusters. This is quite surprising considering the fact that the experimental results for the ionization potential, static polarizability, and photoabsorption spectra of potassium clusters as functions of cluster size were reported very early in the development of cluster physics \cite{knight,clemenger}. Along with the calculations of sodium clusters few SJBM based studies within the realm of DFT and TDDFT  pertaining to the  evolution of binding energies, ionization potentials, static polarizabilities of potassium clusters exist in the literature \cite{kresin1,rubio,balbas,yannouleas,pacheco1,alasia,landmann,pohl}. Likewise, few studies of the potassium clusters at the \textit{ab initio} level taking into account the detailed ionic structure employing  various approaches like configuration interaction (CI) \cite{pacchioni,spiegelmann}, many-body perturbation theory \cite{ray}, DFT coupled with pseudopotential \cite{flad}, and coupled cluster theory \cite{ccsdstoll} have also been reported in the literature . However, these studies were restricted to very small sized clusters containing maximum  up to seven potassium atoms \cite{ray} and investigated ground state properties like bond lengths, binding energies, and ionization potentials. To the best of our knowledge no \textit{ab initio} level calculation of static polarizability either by employing correlated wavefunction approach or by DFT/TDDFT methods exists in the literature. The aim of this paper is to fill this gap by extending the \textit{ab initio} study to larger potassium clusters containing up to 20 potassium atoms and investigate systematically the size evolution of various ground state properties like the binding energy (BE), ionization potential (IP) and response property like static polarizability. For this purpose we employ \textit{ab initio} DFT based method with gradient corrected exchange-correlation (XC) potential for geometry optimization and three methods for the calculations of polarizabilities namely, the second order Moller-Plesset perturbation theory (MP2) \cite{mp2}, coupled cluster theory with single and double excitations and perturbative triple excitations (CCSD(T)) \cite{piecuch},  and TDDFT with asymptotically correct XC potential.  The high computational cost and time involved in CCSD(T) and MP2 based calculations restrict the maximum size of the cluster that could be handled by this approach. In this paper we perform MP2 based polarizability calculations for clusters containing maximum up to 14 atoms and CCSD(T) based calculations could be performed for dimer and tetramer only. Nonetheless these results provide a way to check the correctness and consistencies of the results obtained by us by employing various methods.

We note that to carry out \textit{ab initio} calculations of ground state properties mentioned above and the polarizabilities of clusters  which go beyond jellium model, it is necessary to have the knowledge of the ionic structure of the clusters. To this end, we carry out a systematic search for the optimized structures by employing DFT based geometry optimization scheme with Becke-Perdew (BP86) \cite{beckeperdew} exchange-correlation (XC) potential within the generalized gradient approximation (GGA) as this potential is known to yield reliable geometries. For each cluster beyond K$_{4}$ several structures have been considered as starting geometries. To the best of our knowledge studies involving geometry optimization of potassium clusters beyond K$_{7}$ are not available in the literature and our results in this paper will be useful for further studies on potassium clusters. A brief discussion on geometry optimization procedure is given in next section.   

The calculation of response property like polarizabilities by TDDFT approach requires approximating the forms of XC functionals. It is well known that the accuracy of the results for response properties obtained via TDDFT crucially depends on the nature of the XC potential,  especially its behaviour in the asymptotic region \cite{gisbergen1,banerjee1}. Keeping this in mind, we carry out all-electron TDDFT based calculations of static poarizabilities of potassium clusters with a model XC potential, called statistical average of orbital potentials (SAOP), which has desirable properties both in the asymptotic and the inner regions of a molecule \cite{gritsenko,schipper}. The choice of SAOP is also motivated by the results of Ref. \cite{pacheco3,arup1} where it has been shown that SAOP polarizability and excitation energies of sodium clusters agree well with the experimental data.  In order to study the effect of XC potential on the polarizabilities, calculations of polarizabilities are also carried out with less accurate XC potential under local density approximation (LDA). 

It is well known  that the static polarizabilities of sodium clusters obtained by employing DFT and TDDFT within SJBM are generally underestimated in comparison to the corresponding \textit{ab initio} and experimental results. However, such comparison of the SJBM, \textit{ab initio}, and experimental results does not exist for potassium clusters. In order to test the accuracy of the jellium model, we compare the results for the static polarizability of 8 and 20 atom clusters (as jellium based results only for these two magic clusters are available in the literature) obtained by employing LDA XC potential in the realm of SJBM with corresponding \textit{ab initio} and experimental data.

The rest of the paper is organized as follows: In section II we discuss the theoretical methods employed to calculate the optimized geometry, BE, IP, and static polarizabilities of potassium clusters. Results of our calculations are presented in Section III and the paper is concluded in Section IV.

\section{Theoretical Methods}
In this work we study the structure and other properties of small sized potassium clusters on the basis of all-electron \textit{ab initio} methods. For this purpose we use ADF program package \cite{adf} for DFT and TDDFT based calculations and GAMESS electronic structure code for carrying out post-Hartree-Fock MP2  and CCSD(T) calculations \cite{gamess}. We calculate the optimized geometries of clusters containing even number of atoms ranging from 2 to 20 atoms. The geometry optimizations of all the clusters have been performed through DFT based calculations by employing a triple-$\xi$ Slater-type orbital (STO) basis set with two added polarization functions (TZ2P basis set of ADF basis set library) along with Becke-Perdew (BP86) XC potential \cite{beckeperdew}. The geometry optimization involves finding local minima on the multidimensional potential energy surface. The staring geometry of the cluster plays an important role in the optimization procedure. In the present calculations for each cluster we have considered more than one starting geometries which are compiled from already available optimized structures of sodium clusters \cite{solovyov,kronik}.  Here we make no assumption regarding the core electrons and perform all-electron  geometry optimization
calculations. All the optimizations are carried out with the convergence criteria for the norm of energy gradient and energy, fixed at $10^{-4}$ atomic units (a.u.) and $10^{-6}$ a.u., respectively. 

In this paper we further calculate the static polarizabilities of potassium clusters by employing MP2, CCSD(T), and  TDDFT based methods. The calculations of polarizabilities with MP2 and CCSD(T) have been carried out by employing finite field approach available in the GAMESS electronic structure code. The finite field approach makes use of the perturbative series expansion of the energy $E$ in terms of the components of a static uniform electric field $\vec{F}$  given by,
\begin{equation}
E(\vec{F}) = E(0) + \sum_{i}\mu_{i}F_{i} + \frac{1}{2}\sum_{ij}\alpha_{ij}F_{i}F_{j} + \cdots,
\end{equation}
where $E(0)$ is the energy of the system in the absence of the applied electric field, $\vec{\mu}$ is dipole moment, and $\alpha_{ij}$ $( i,j = x, y, z)$ is the dipole polarizability tensor. The components of the polarizabilty tensor are obtained as the second-order derivatives of the energy with respect to the components of the electric field,
\begin{equation}
\alpha_{ij} = \left ( \frac{d^{2}E}{dF_{i}dF_{j}}\right)_{F = 0}.
\end{equation}
The derivatives are calculated numerically by applying fields of  0, 0.001 and 0.002 a.u. along $\pm x$, $\pm y$, and $\pm z$ directions and mean polarizability is calculated from the diagonal elements of polarizability tensor as
\begin{equation}
\bar{\alpha} = \frac{1}{3}\left (\alpha_{xx} + \alpha_{yy} + \alpha_{zz}\right ).
\end{equation} 
All MP2 and CCSD(T) calculations have been performed with  valence triple zeta polarized Gaussian basis sets of Sadlej and Urban \cite{sadlej} for potassium atom.

On the other hand, TDDFT calculation of polarizability is based on the linear response theory of many-body systems and employs exact analytical expressions for polarizability in terms of the moment of the first-order induced density. To avoid digression we refer the readers to Ref. \cite{gisbergen} for detailed description of the linear response theory based method which is adopted in ADF program package for obtaining polarizability. We note here that TDDFT based method gives frequency dependent polarizability but here we focus our attention on the static or zero-frequency polarizability. We also mention here that the calculation of polarizability based on linear response theory of many-body system yields results which are more accurate than the finite field approach as no explicit specification of the magnitude of the applied field is required in the former method. 
  It has been already mentioned that TDDFT based response property calculation requires approximating the XC functional at two different levels. The first one is the static XC potential needed to calculate the ground-state Kohn-Sham (KS) orbitals and orbital energies. The second approximation is needed to represent the XC kernel $f_{XC}({\bf r},{\bf r'},\omega)$ which determines the XC contribution to the screening of an applied field. For the XC kernel, we use reasonably accurate adiabatic local density approximation (ALDA) \cite{petersilka}. On the other hand, for the static XC potential needed to calculate the  ground-state orbitals and energies, two different choices have been made.  These are (i) the standard potential under local density approximation (LDA) as parametrized by Vosko, Wilk and Nussair \cite{vwn} and (ii) the model potential SAOP possessing correct behaviour both inner and asymptotic regions \cite{gritsenko,schipper}. The results obtained by these two XC potentials are compared in order to investigate the effect of XC potential on the results for the polarizability. 
The calculations of polarizabilities of potassium clusters by TDDFT based method are carried out by using large Slater type orbital (STO) basis sets. It is well known that for accurate calculations of response properties it is necessary to have large basis sets with both polarization and diffuse functions. For our purpose, we have chosen one of the largest all electron even tempered basis set ET-QZ3P-2DIFFUSE with two sets of diffuse functions consisting of (13s,10p,5d,3f) functions for K available in the ADF basis set library.  The application of basis set with diffuse functions often leads to the problem of linear dependencies. Such problems have been circumvented by removing linear combinations of functions corresponding to small eigenvalues of the overlap matrix. We expect that the size of the chosen basis set will make our results very close to the basis-set limit. The next section is devoted to the discussion of the results for the structures and properties of potassium clusters.

\section{Results and Discussion}
We begin this section with the discussion on the results of our geometry optimization calculations followed by the results for BE, IP, and polarizabilities of clusters consisting of even number of atoms ranging from 2 to 20. We compare the results of our calculations with the available experimental data and also results of other theoretical works performed both within the framework of the jellium model and beyond, using DFT or correlated wavefunction based methods, and assess the level of accuracy of different theoretical approaches. 
\subsection{Structure and other properties of K$_{n}$ clusters}
The optimized structures of potassium clusters with even number of atoms up 
to 20 are shown in Fig. 1 ($n=$2, 4, and 6) and Fig. 2 ($n\geq 8$). The 
indices $n$ and $m$ in the label $n\_m$ assigned to each cluster in these 
figures denote the number of atoms and the rank in the increasing energy order.
 In order to perform the geometry optimization calculations of potassium clusters, we make use of the optimized geometries of sodium clusters reported in Refs. \cite{kronik,solovyov}, as initial geometries. For each cluster, we consider all the isomers available in Refs. \cite{kronik,solovyov} to search for the minimum energy structures. We note here that structures reported in these two papers for sodium clusters are quite exhaustive and include structures obtained by several other authors also. Therefore we feel that the minimum energy structures obtained by us represent true ground state structures of the potassium clusters.  Now we present the results for BE per atom and the average inter-atomic distances for all the isomers for each cluster. We tabulate the results in two parts. In Table I, we present the results for K$_{2}$, K$_{4}$, and K$_{6}$ clusters and compare them with the other theoretical \cite{pacchioni,spiegelmann,ray,flad} and experimental (for dimer) \cite{huber} data available in literature. On the other hand, the results for the K$_{8}$ and higher clusters are presented in Table II 
for which no data are avaiable for comparison. We calculate BE which is
 presented in Table I and II by employing the formula

\begin{equation}
E_{b}(K_{n}) = nE(K) - E(K_{n})
\label{bindingenergy}
\end{equation}
where $E(K_{n})$ and $E(K)$ are the total energies of a neutral n-atom potassium cluster $K_{n}$ and an isolated single potassium atom, respectively. Note that according to Eq.\ref{bindingenergy}, BE is a negative number for a bound structure and larger value of it implies a more stable structure. The second quantity for which results are presented in Table I and II is the average inter-atomic distance in each isomer of a cluster. This quantity is computed by employing the corresponding optimized structure and considering only inter-atomic distances smaller than 5.076$\AA$ which is 10$\%$ higher than the nearest neighbour disatance
in the bcc lattice of bulk potassium.

First we discuss the results for small clusters K$_{2}$, K$_{4}$, and K$_{6}$ (Table I) as for these systems we can assess the accuracy of our DFT based results by comparing them with the published data that already exist in the literature for these three clusters. In Refs. \cite{pacchioni,spiegelmann,flad,ccsdstoll} the calculations on above mentioned clusters were carried out by employing pseudopotential method in conjunction with configuration interaction (CI) approach \cite{pacchioni,spiegelmann}, self interaction corrected DFT \cite{flad} method, and CCSD(T) for the valence electrons \cite{ccsdstoll}. On the other hand, an all-electron calculation using the techniques of Hartree-Fock theory followed by many-body perturbation theory (MBPT) was carried out to determine the equilibrium geometries of potassium clusters up to $K_{7}$ \cite{ray}.  It can be seen from Table I that for dimer, results for both BE and average interatomic distance obtained by us employing DFT with BP86 XC potential are quite close to the other published data including the experimental result \cite{huber}. In fact our result for the BE per atom for $K_{2}$ matches exactly with the experimental data and our result for interatomic distance is closest to the experimental value compared to other theoretical data. 

Following previous studies \cite{pacchioni,spiegelmann,flad}, we consider rhomboidal geometry for tetramer $K_{4}$.  The average interatomic distance for the tetramer geometry obtained by us is quite close to other results and same is true for the BE per atom except for the result of Ref. \cite{pacchioni}. For $K_{6}$, two geometries, namely a planar and a three-dimensional structures have been considered for the geometry optimization. We find that the planar isomer is higher in energy than the three-dimensional one in conformity with the results of Refs. \cite{spiegelmann,ray}. For the minimum energy structure of $K_{6}$ the difference between the results for BE and average interatomic distance obtained by us and those of Refs. \cite{spiegelmann,ray} is less than 10$\%$.  As mentioned before for clusters beyond K$_{7}$ no results for the optimized geometries of potassium clusters are available in the literature for guidance. Consequently, to obtain the geometries of potassium clusters containing 8 and more atoms we make use of corresponding geometries of sodium clusters reported in Refs. \cite{kronik,solovyov} as the starting geometries and optimize them by employing DFT based method with TZ2P basis set and BP86 XC potential. For K$_{8}$ we consider two stuructures one possessing  D$_{2d}$ symmetry \cite{kronik} and another having T$_{d}$ symmetry \cite{solovyov}. We find that D$_{2d}$ symmetry isomer of K$_{8}$ is lower in energy than T$_{d}$ structure by around 0.08 eV. 

For clusters  beyond Na$_{8}$, many (more than two) quasi-degenerate isomers exist. We use all these isomers for the geometry optimization of potassium clusters. In order to perform the geometry optimization calculations for the clusters beyond K$_{8}$, we sample four isomers for K$_{10}$, three isomers each for K$_{12}$, K$_{14}$, K$_{16}$ and K$_{18}$ and five isomers for K$_{20}$. These optimized structures in increasing energy order are shown in Fig. 2 and their BE and average interatomic distances are presented in Table II. We note here that for K$_{20}$ the higher symmetry  T$_{d}$ isomer does not yield minimum energy    
and it is for a lower symmetry D$_{2d}$ isomer of 20 atom cluster we get minimum energy. The energy of T$_{d}$ isomer is around 0.37 eV higher than the minimum energy D$_{2d}$ structure. It is interesting to note that for the two magic number clusters K$_{8}$ and K$_{20}$, the minimum energy geometries possess D$_{2d}$ symmetry.

Having determined the geometry of potassium clusters containing even number of atoms up to K$_{20}$, we next focus our attention on the evolution of ionization potential (IP) with the size of clusters, as this dependence has been investigated extensively and these results are available in the literature \cite{clemenger,herman,brechignacip,kappes}. Apart from these experimental results some papers also reported theoretical results for IP of small potassium clusters at the \textit{ab initio} level \cite{spiegelmann,flad} and also within SBJM \cite{balbas}. We compare these results with the ones obtained in this paper by DFT based calculations with TZ2P basis set and BP86 XC potential. To calculate IP, we restrict ourselves to only the minimum energy isomers for each cluster and employ following formula of IP of a cluster containing N atoms:
\begin{equation}
IP = E(K_{n})^{+} - E(K_{n}),
\label{ionizatioppotential}
\end{equation}
where E($K_{n})^{+}$ and E($K_{n}$) are energies of the singly charged and neutral clusters respectively. 
The results of these calculations along with the data available in the literature are presented in Table III and Fig. 3.  In Table III, we  present the results for small sized potassium clusters containing 2, 4, and 6 atoms for which both theoretical and experimental results are available. For clusters containing more than 6 atoms theoretical results at \textit{ab initio} level are not available, so we compare our results with the experimental data only and these are displayed in Fig. 3. From Table III, it can be seen that for K$_{2}$, K$_{4}$, and K$_{6}$, our DFT based results for IP are quite close to other theoretical as well as experimental data. Fig. 3 shows the dependence of cluster IP obtained by us on $n$ along with the experimental results of Ref. \cite{clemenger,herman,brechignacip,kappes}. This comparison clearly shows that our results follow a similar trend as compared to the experimental data with increase in $n$. However, the theoretical results are slightly overestimated with respect to the experimental ones. This may be attributed to
the fact that the experiments have been performed at finite temperatures \cite{landmann}.
Here we wish to point out that Fig. 3 also clearly elucidates that IP of potassium clusters decreases with increasing cluster size, which is in conformity with the conducting sphere model (CSM) of metal cluster \cite{cini,wood,perdewip}. To verify this we calculate IP in accordance with the expression
\begin{equation}
IP  =  W + \frac{3}{8}\frac{e^{2}}{R} 
\label{ipconductingsphere}
\end{equation}
where W is the bulk work function and R is radius of the metallic droplet. Following Ref. \cite{brechignacip1}, we use W = 2.28 eV and R $=r_{s}n^{1/3}$ ( where $=r_{s}$ is the Wigner-Seitz radius) with $r_{s} = 4.86 a.u.$. These results are shown in Fig. 3 by dashed line and our DFT result follow a similar trend.  
\subsection{Polarizability of potassium clusters}
In this section we present and discuss the results of our calculations for the static dipole polarizabilities of potassium clusters. The static polarizability plays an important role in the charctarization of the clusters and it is one of the property which has been extensively  measured for metal clusters specially sodium clusters \cite{knight,brechignac,rayane,tikhonov}. To the best of our knowledge the experimental results for static polarizabilities of some potassium clusters are available only in Ref. \cite{knight}. Here we will compare results of our calculations for $\bar{\alpha}$ of clusters containing 2, 8, and 20 atoms with the above-mentioned experimental results.  

We begin our discussion on the results for polarizabilities  by first testing the accuracies of SAOP and MP2 results for dimer and tetramer against the corresponding data obtained with a large basis set coupled cluster theory with single and double excitations and perturbative triple excitation (CCSD(T)) calculations \cite{piecuch}.
In Table IV we present the results for polarizabilities of K$_{2}$ and K$_{4}$ obtained with three different methods along with the other theoretical \cite{urban,ccsdstoll} and experimental \cite{clemenger,tarnovsky} results for the dimer which are already available in literature. From Table IV, we notice that the results for the dimer obtained by CCSD(T) approach vary from 486 to 510 a.u. and these results are well within the experimental errors. The variation in the different CCSD(T) results is attributed to the use of different basis sets and also to the level of calculations (all-electron or pesudopotential). The basis set used in the present paper is similar to the one employed in Ref. \cite{urban} and both are all-electron calculations and consequently two results are close to each other. The difference in the two results may be due to the use of slightly different bond length ($3.91 \AA$) of the dimer  and also inclusion of relativistic effect in the calculation of Ref. \cite{urban}. Furthermore, we observe that MP2 value for the polarizability of K$_{2}$ is quite close to the lowest CCSD(T) result of Ref. \cite{ccsdstoll}, however, SAOP underestimates the polarizability by around $6\%$ with respect to CCSD(T) rseults. 
In contrast to the dimer case SAOP polarizability for K$_{4}$ is quite close to the CCSD(T) result and slightly higher (around 2$\%$) than MP2 result.  

We now proceed with the calculations of 
static polarizabilities for all the optimized isomers shown in Figs. 1 and 2. We note here that the geometries considered in this paper are nonspherical and, consequently the polarizability tensors are expected to be anisotropic. We also calculate the anisotropy in polarizability given by 
\begin{equation}
|\Delta\alpha|  =  \left [\frac{3Tr{\bf\alpha}^{2} - \left (Tr{\bf\alpha}\right )^{2}}{2}\right ]^{1/2}  \qquad {\rm (general}\quad
{\rm axes})\qquad
\end{equation}
where $\bf{\alpha}$ is the second-rank polarizability tensor.

The two methods (MP2, and TDDFT), which we employ to calculate the static polarizabilities of clusters (for $n > 4$) take into account the electron correlations in different ways and thereby enabling us to check the consistencies of our results, as no other theoretical results are available in the literature for comparison. In Table V we present results for the average static polarizability $\bar{\alpha}$ and anisotropy in the polarizability $\Delta\alpha$ obtained with MP2 and TDDFT (with SAOP and LDA XC potentials) methods.  We have performed MP2 calculations of $\bar{\alpha}$ and $\Delta\alpha$ for clusters only up to K$_{14}$ and K$_{10}$, respectively. We note from Table V that the results for both polarizabilities and their anisotropies obtained by MP2 and TDDFT-SAOP are quite close for all the clusters while corresponding LDA values are systematically underestimated. SAOP results for all the isomers for each cluster are actually slightly higher than the corresponding MP2 data except for K$_{2}$. In order to study the evolution of polarizability $\bar{\alpha}$ and anisotropy in polarizability $\Delta\alpha$ with the size, we plot them in Figs. 4a  and 4b,  respectively for the minimum energy isomers of each cluster as a function of number of atoms. In this figure we also display experimental data for three magic clusters, namely, K$_{2}$, K$_{8}$, and K$_{20}$ \cite{knight}. Fig. 4 once again clearly shows that MP2 results for the polarizabilites and anisotropies in polarizabilities are very close to the corresponding SAOP results and LDA values are systematically lower than  both of them.  Both SAOP and MP2 results for the polarizabilities for magic number clusters K$_{2}$, K$_{8}$ and K$_{20}$ are well within the experimental error bars (of the order of
$\pm$ $5-7\%$). The maximum difference between the experimental and SAOP as well as MP2 results is observed for K$_{20}$ cluster. Note also (from Table V) that both SAOP and MP2 results for the polarizabilities of clusters K$_{8}$ and K$_{20}$ possessing T$_{d}$ symmetry ( K$_{8\_1}$ and K$_{20\_4}$ ) are closer to the corresponding experimental data (for K$_{8}$, $\bar{\alpha} = 1653 \pm 83$ a.u. and 
for K$_{20}$ $\bar{\alpha} = 3834 \pm 300$ a.u. from experiments) than their respective minimum energy structures (see Table V).  From these results it may be inferred that for these two clusters the geometries detected in the experiments performed at finite
temperatures may be different from what have been obtained in this paper by DFT based calculations.   Overall, we conclude from the results of Table V and Fig. 3 that TDDFT based calculations with SAOP yield results for polarizability which are quite accurate and compare well with the correlated wavefunction based MP2
results as well as experimental data. 

It has already been pointed out that MP2 calculations are computationally expensive and thus it becomes increasingly difficult to apply this method for very large clusters (more than 10 atoms). 
However, a good match between SAOP and MP2 results for clusters up to K$_{14}$ encourages us to explore how the two results scale with respect to each other. To this end we plot SAOP and MP2 results along y- and x-axes, respectively, in Fig. 5 and fit the data points with a straight line by least square fitting. It can be clearly seen from Fig. 5 that a very good fitting is obtained with the correlation coefficient value 0.9997 signifying a linear relationship between SAOP and MP2 results. This linear relationship between SAOP and MP2 will enable us to predict MP2 results for the polarizability of larger clusters. 

According to the jellium model, clusters with closed shells of delocalized electrons have spherical shape \cite{brackreview,alonso}. It is well known that polarizability of such a sphere is proportional to the volume of the sphere.  Since the geometries of clusters considered in this paper are, in general, not spherical in nature and thus such linear dependence of polarizabiliy with the volume of cluster is not very obvious. However, the studies on the relationship between the static polarizability and the volume of carbon and sodium clusters have already been reported in the literature \cite{ghanty1,arup1,ghantyhard} . Here we extend this study for potassium clusters and we go up to cluster containing 20 atoms. For this purpose we use SAOP results for the polarizabilities of lowest energy isomers for each cluster and obtain its volume by using the prescription of Tomasi and Persico \cite{tomasi}. The plot of the polarizability as a function of the volume of the clusters is shown in Fig. 6. It can be clearly seen from this figure that a good fitting is obtained with the correlation coefficient value of 0.996. This result then clearly suggests that  a good correlation exists between the polarizability and the cluster volume even for nonspherical potassium clusters. This linear correlation between the polarizability and volume is an important result as it enables us to construct a size-to-property relationship for polarizability.  Using this relationship polarizabilities of larger clusters can be calculated as for these clusters performing \textit{ab initio} calculations are computationally expensive if not impossible. 

We wish to close this paper with a comparison of these SJBM results with the \textit{ab initio} results presented in this paper. The jellium based results within DFT for the polarizability of K$_{8}$, and K$_{20}$ are available in the literature \cite{rubio} and we compare them  with corresponding \textit{ab initio} TDDFT results. The LDA calculations in Ref. \cite{rubio} were performed by employing Dirac form  for the exchange and Wigner form for the correlation energy functionals \cite{diracwigner}. On the other hand, in this paper we employ VWN parametrization of the LDA XC functional
which uses same Dirac exchange energy functional, but the parametrization for correlation part is different from the Wigner functional. We expect that the deviation in the results due to application of different correlation energy functionals will be significantly smaller than the difference in the two results, arising due to the consideration of structures of the clusters in \textit{ab initio} calculations. The results for the polarizabilities of K$_{8}$ and K$_{20}$ clusters calculated with LDA XC functional in jellium model are found to be $\alpha_{K_{8}} = 1212$ a.u. and $\alpha_{K_{20}} = 2939$ a.u.. In comparison to this our \textit{ab initio} results (for the minimum energy geometries) with LDA XC potential are $\alpha_{K_{8}} = 1367$ a.u. and $\alpha_{K_{20}} = 3220$ a.u.. Besides LDA, Rubio et al. \cite{rubio} also employed a potential with correct $-1/r$ asymptotic decay as introduced by Przybylski and Borstel (PB) \cite{przybylski} within weighed density approximation (WDA) to calculate the polarizability. We compare the results obtained with PB potential within jellium model ( $\alpha_{K_{8}} = 1542$ a.u. and $\alpha_{K_{20}} = 3489$ a.u.) with SAOP results ($\alpha_{K_{8}} = 1531$ a.u. and $\alpha_{K_{20}} = 3582$ a.u.) as SAOP too possesses correct asymptotic behaviour.  The jellium based LDA results are around 10$\%$ lower than the corresponding TDDFT-based \textit{ab initio} values. On the other hand, differences between \textit{ab initio} SAOP and PB within jellium model results are further smaller. We note here that a similar observation for the polarizabilities of sodium clusters was made in Ref. \cite{solovyov}.
From the closeness of the results obtained by employing \textit{ab initio} and SJBM, we conclude that for the alkali-metal atom clusters detailed ionic core structures may not have much influence on the values of the cluster polarizabilties.   
\section{Conclusion}
This paper is devoted to the calculations of the optimized geometries and various other properties of potassium clusters containing even number of atoms ranging from 2 to 20. In order to determine optimized geometry we have used DFT with TZ2P basis set and GGA XC potential. For each cluster beyond K$_{4}$ more than one geometry has been considered to explore the possibility of the existence of various structural isomers. To accomplish this we have taken various optimized geometries of sodium clusters available in the literature as our starting geometries. For small clusters containing 2, 4, and 6 atoms the results of our calculations for BE per atom, bond length, and average interatomic distance match quite well with the other published data obtained with correlated wavefunction based methods. 

We have also studied from these DFT based calculations evolution of IP with the size of clusters. The experimental data for the size dependence of IP of potassium clusters is available in the literature and our \textit{ab initio} DFT based results match quite well with them. 

In this paper we have carried out calculations of the static polarizabiliteis of the potassium clusters with MP2 and TDDFT approaches as well as CCSD(T) (for dimer and tetramer) taking electron correlations in different ways. A model XC potential (SAOP) possessing correct behaviours both in the asymptotic and inner regions of the molecule and also less accurate LDA XC potential have been used to calculate polarizabilities within TDDFT. For all the calculations sufficiently large basis sets have been employed. For dimer and tetramer the results for the polarizabilities obtained by different methods, employed in this paper, agree witheach other. Similarly for clusters beyond K$_{4}$ and up to K$_{14}$ MP2 and SAOP results for the polarizabilities are quite close to each other. We find a very good linear correlation between MP2 and SAOP results. On the other hand, TDDFT based calculations with LDA XC potentials are systematically lower than those of MP2 and SAOP. Moreover, we also find that both SAOP and MP2 results for the static polarizabilites of 2-, 8-, and 20-atom potassium clusters agree quite well with the experimental results. In general it is observed that both for sodium and potassium clusters the SAOP data for the polarizability is higher than the corresponding MP2 results. In this paper we have also investigated the volume-to-polarizability scaling for the potassium clusters. Our study has found a very good linear correlation between the volume and polarizability of the clusters. This scaling law can be exploited to determine the polarizabilities of larger clusters. Finally we have also compared the SJBM based results for the polarizabilities of K$_{8}$ and K$_{20}$ with our corresponding \textit{ab initio} values obtained by employing TDDFT. These comparison clearly reveals that jellium model based results for the polarizabilities are quite accurate for magic number clusters and it is expected that this model is increasingly more suitable for such larger clusters.

\acknowledgments{ A. B. and A. C. wish to thank Pranabesh Thander of
RRCAT Computer Centre for his help and support in providing us
uninterrupted computational resources and also for smooth running
of the codes and C. Kamal for his help in geometry optimization. 
It is a pleasure to thank Prof. Vitaly Kresin
for his valuable suggestions and making some experimental data
available to us.}

\clearpage
\newpage
\section*{Figure captions}
{\bf Fig.1} Optimized ground state geometries of K$_{2}$, K$_{4}$, and K$_{6}$ clusters. Letters denote the dimensions tabulated in Table I.

{\bf Fig.2} Optimized geometries of potassium clusters K$_{n}$ with $n = 8-20$. Binding energy per atom and average interatomic distance for these clusters are tabulated in Table II.

{\bf Fig.3} Plot of ionization potential (in eV) of potassium clusters containing even number of atoms ranging from 2 to 20 atoms as a function of number of atoms. The results obtained with BP86 XC potential ( solid squares) are compared with the experimental results (solid circles) and (solid triangles) of Refs. \cite{clemenger} and \cite{kappes} respectively. The continuous dotted line shows the results obtained via Eq. \ref{ipconductingsphere}.

{\bf Fig.4} Plot of average static polarizability (a) $\bar{\alpha}$  and (b) anisotropy in polarizability $\Delta\alpha$ for minimum energy isomers of potassium clusters as a function of number of particles. The experimental results for the polarizabilities of magic clusters K$_{2}$, K$_{8}$, and K$_{20}$ \cite{knight} are also shown in this figure. The lines joining the points are guide to the eye.

{\bf Fig.5} Plot of SAOP results for average static polarizability $\bar{\alpha}$ against corresponding MP2 values.  All the results are in atomic units and straight line is least square fitted line.  

{\bf Fig.6} Plot of average static polarizability $\bar{\alpha}$ obtained with SAOP as a function of the cluster volume. All the results are in atomic units and straight line is least square fitted line.  
\newpage
\begin{table}
\caption{  Comparison of binding energy per atom ( in eV), bond length, and average interatomic distance $\langle R\rangle$ (in angstrom) for K$_{2}$, K$_{4}$, and K$_{6}$ clusters.}  \tabcolsep=0.1in
\begin{center}
\begin{tabular}{c|c|c|c|c|}
K$_{n\_m}$& Reference  & -BE/N  & Bond length & $\langle R\rangle$ \\
\hline
K$_{2\_0}$ & Present & 0.25 & 3.94 & 3.94   \\
   & Ref. \cite{ray} &0.19 & 4.22 & 4.22  \\
   & Ref. \cite{pacchioni}& 0.21 & 4.21 & 4.21  \\
   &Ref. \cite{spiegelmann} & 0.32 & 3.84 &3.84 \\
   &Ref. \cite{flad}& 0.26 & 4.05 & 4.05  \\
   &Ref. \cite{ccsdstoll}& 0.27 & 3.92 & 3.92 \\
   &Ref. \cite{huber}(Expt.) & 0.25 & 3.90 & 3.90  \\
\hline  
K$_{4\_0}$ & Present & 0.31 & a$=4.44$, b$ = 3.98$ & 4.34  \\ 
      & Ref. \cite{ray} &0.28 & $a = b = 4.78$ & 4.78  \\
      &Ref. \cite{pacchioni}& 0.19 & $a = b = 4.90 $ & 4.90  \\
      &Ref. \cite{spiegelmann} & 0.37 & $a = b = 4.44 $ & 4.44  \\
      &Ref. \cite{flad}& 0.34 & a$=4.42$, b$ = 3.92$ & 4.44  \\
\hline  
K$_{6\_0}$ & Present & 0.39 & a$=4.47$, b$ = 4.31$ & 4.39  \\ 
         & Ref. \cite{ray} &0.37 & $ a = b = 4.65$ & 4.65  \\
         &Ref. \cite{spiegelmann} & 0.44 & $a = b = 4.50 $ & 4.50  \\
K$_{6\_1}$ & Present & 0.39 & a$=4.29$, b$ = 4.60$ & 4.39  \\ 
        & Ref. \cite{ray} &0.36 & $ a = b = 4.65 $ & 4.65  \\
         &Ref. \cite{spiegelmann} & 0.42 & $a = b = 4.28$ & 4.28  \\       
\end{tabular}
\end{center}
\end{table}

\begin{table}
\caption{ Binding energy per atom ( in eV), and  average interatomic distance $\langle R\rangle$ (in angstrom) of potassium clusters containing 8 to 20 atoms.} \tabcolsep=0.5in
\begin{center}
\begin{tabular}{c|c|c|}
K$_{n\_m}$ &  -BE/N  & $\langle R\rangle$ \\
\hline
K$_{8\_0}$ & 0.458 & 4.44   \\
K$_{8\_1}$ & 0.448 & 4.45  \\
\hline  
K$_{10\_0}$ & 0.457 & 4.48 \\
K$_{10\_1}$ & 0.454 & 4.50 \\
K$_{10\_2}$ & 0.453 & 4.54 \\
K$_{10\_3}$ & 0.413 & 4.50 \\
\hline  
K$_{12\_0}$ & 0.476 & 4.54 \\
K$_{12\_1}$ & 0.473 & 4.50 \\
K$_{12\_2}$ & 0.471 & 4.51 \\
\hline  
K$_{14\_0}$ & 0.493 & 4.57 \\
K$_{14\_1}$ & 0.487 & 4.50 \\
K$_{14\_2}$ & 0.486 & 4.48 \\ 
\hline  
K$_{16\_0}$ & 0.500 & 4.57 \\
K$_{16\_1}$ & 0.499 & 4.57 \\
K$_{16\_2}$ & 0.498 & 4.53 \\  
\hline  
K$_{18\_0}$ & 0.522& 4.52 \\
K$_{18\_1}$ & 0.518 & 4.60 \\
K$_{18\_2}$ & 0.515 & 4.62 \\
\hline  
K$_{20\_0}$ & 0.533 & 4.53 \\
K$_{20\_1}$ & 0.532 & 4.57 \\
K$_{20\_2}$ & 0.530 & 4.59 \\
K$_{20\_3}$ & 0.529 & 4.52 \\
K$_{20\_4}$ & 0.514 & 4.56 \\       
\end{tabular}
\end{center}
\end{table}

\begin{table}
\caption{ Comparison of the theoretical and experimental IP (in eV) for K$_{2}$, K$_{4}$, and K$_{6}$ clusters.} \tabcolsep=0.25in
\begin{center}
\begin{tabular}{c|c|c|c|}
Reference & K$_{2}$ & K$_{4}$ & K$_{6}$  \\
\hline
Present (BP86) & 4.28 & 3.61 & 3.73  \\
Ref. \cite{pacchioni}& 3.79 & 3.23 &  \\
Ref. \cite{spiegelmann} & 4.12 & 3.64 & 3.79  \\
Ref. \cite{flad}& 4.32 & 3.43 &  \\
Ref. \cite{clemenger}(Expt.) & 4.05 & 3.52 & 3.35  \\ 
Ref. \cite{kappes}(Expt.) & 4.05$\pm$ 0.05 & 3.6$\pm$ 0.1 & 3.44$\pm$ 0.1 \\ 
Ref. \cite{herman}(Expt.) & 4.05$\pm$ 0.05 & 3.6$\pm$ 0.1 &   \\
Ref. \cite{brechignacip}(Expt.) &  & 3.36 & 3.25 \\
   
\end{tabular}
\end{center}
\end{table}

\begin{table}
\caption{ Comparison of the average static polarizabilty $\bar{\alpha}$ (in a.u.) for K$_{2}$ and K$_{4}$, clusters.} \tabcolsep=0.25in
\begin{center}
\begin{tabular}{c|c|c|}
Method (Reference) & K$_{2}$ & K$_{4}$  \\
\hline
SAOP (Present) & 471.2 & 994.1  \\
MP2 (Present) & 483.2 & 977.8  \\
CCSD(T) (Present) & 510.2 & 991.0  \\
CCSD(T) (Ref. \cite{ccsdstoll}) & 486.4 &  \\
CCSD(T) (Ref. \cite{urban}) & 502.1 &  \\
Expt. (Ref. \cite{clemenger}) & 486.5 &  \\ 
Expt. (Ref. \cite{tarnovsky}) & 500$\pm 40$ &  \\
\end{tabular}
\end{center}
\end{table}

\begin{table}
\caption{Average static polarizability $\bar{\alpha}$ and anisotropy in polarizability $\Delta\alpha$ of various isomers of potassium clusters in atomic units.}
\tabcolsep=0.25in
{\begin{tabular}{@{}|c|c|c|c|c|c|c|@{}}\toprule
 & \multicolumn{2}{c|}{MP2} & \multicolumn{2}{c|}{SAOP} & \multicolumn{2}{c|}{LDA} \\
\cline{2-7}
K$_{n}$ & $\bar{\alpha}$ & $\Delta\alpha$  & $\bar{\alpha}$ & $\Delta\alpha$ & $\bar{\alpha}$ & $\Delta\alpha$   \\
\hline
K$_{2}$ & 483.2 & 369.1 & 471.02 & 336.9 & 437.4 & 289.7 \\
\hline
K$_{4}$ & 977.8 & 931.1 & 994.1 & 937.5 & 907.3 & 876.5 \\
\hline
K$_{6\_0}$ & 1308.3 & 754.09 & 1321.1 & 783.22 & 1203.1 & 722.5 \\ 
K$_{6\_1}$ & 1413.8 & 977.5 & 1440.0 & 1011.9 & 1330.0 & 947.1 \\ 
\hline
K$_{8\_0}$ & 1492.2 & 301.0 & 1531.2 & 320.24 & 1367.5 & 267.9 \\ 
K$_{8\_1}$ & 1626.9 & 20.6  & 1649 & 21.85 & 1489.4 & 17.45 \\ 
\hline
K$_{10\_0}$ & 2006.7 & 1204.3 & 2052.6 & 1178.9 & 1849.4 & 1125.7 \\ 
K$_{10\_1}$ & 2013.4 & 1287.2  & 2052.8 & 1288.1 & 1849.8 & 1221.1 \\ 
K$_{10\_2}$ & 2054.0 &  & 2085.6 & 1161.0 & 1878.6 & 1115.1 \\ 
K$_{10\_3}$ & 2558.4 & 68.5  & 2420.5 & 90.70 & 2238.8 & 97.7 \\ 
\hline
K$_{12\_0}$ & 2385.4 &  & 2429.4 & 1395.4 & 2194.6 & 1315.7 \\ 
K$_{12\_1}$ & 2448.2 &   & 2483.1 & 1461.8 & 2259.6 & 1390.5 \\ 
K$_{12\_2}$ & 2380.7 & & 2417.9 & 1467.8 & 2186.2 & 1373.6 \\ 
\hline
K$_{14\_0}$ & 2781.6 &  & 2782.3 & 1516.3 & 2536.4 & 1434.6 \\ 
K$_{14\_1}$ & 2766.6 &   & 2762.3 & 1536.9 & 2517.9 & 1448.9 \\ 
K$_{14\_2}$ & 2661.6 &   & 2733.9 & 1255.6 & 2472.8 & 1205.9 \\ 

\hline
K$_{16\_0}$ &  &  & 2883.2 & 825.4 & 2619.6 & 778.6 \\ 
K$_{16\_1}$ &  &   & 2878.9 & 816.9 & 2614.5 & 769.4 \\ 
K$_{16\_2}$ &  &  & 2990.9 & 912.4 & 2722.3 & 878.4 \\
\hline
K$_{18\_0}$ &  & & 3048.7 & 483.1& 2787.7 & 456.0 \\ 
K$_{18\_1}$ &  &   & 3216.3 & 623.7 & 2912.5 & 568.0 \\ 
K$_{18\_2}$ & &  & 3197.7 & 801.9 & 2894.0 & 744.4 \\  
\hline
K$_{20\_0}$ &  &  & 3582.2& 344.40 & 3220.4 & 184.4 \\ 
K$_{20\_1}$ &  &   & 3604.3 & 140.1 & 3289.1 & 120.8 \\ 
K$_{20\_2}$ &  &  & 3526.4& 539.3 & 3203.0 & 475.2 \\
K$_{20\_3}$ &  &   & 3518.6 & 204.4 & 3220.4 & 184.8 \\ 
K$_{20\_4}$ &  &  & 4066.1& 6.35 & 3769.8& 6.58 \\
\hline
\end{tabular}}
\end{table}

\newpage
\clearpage
\begin{figure}
\begin{center}
\includegraphics{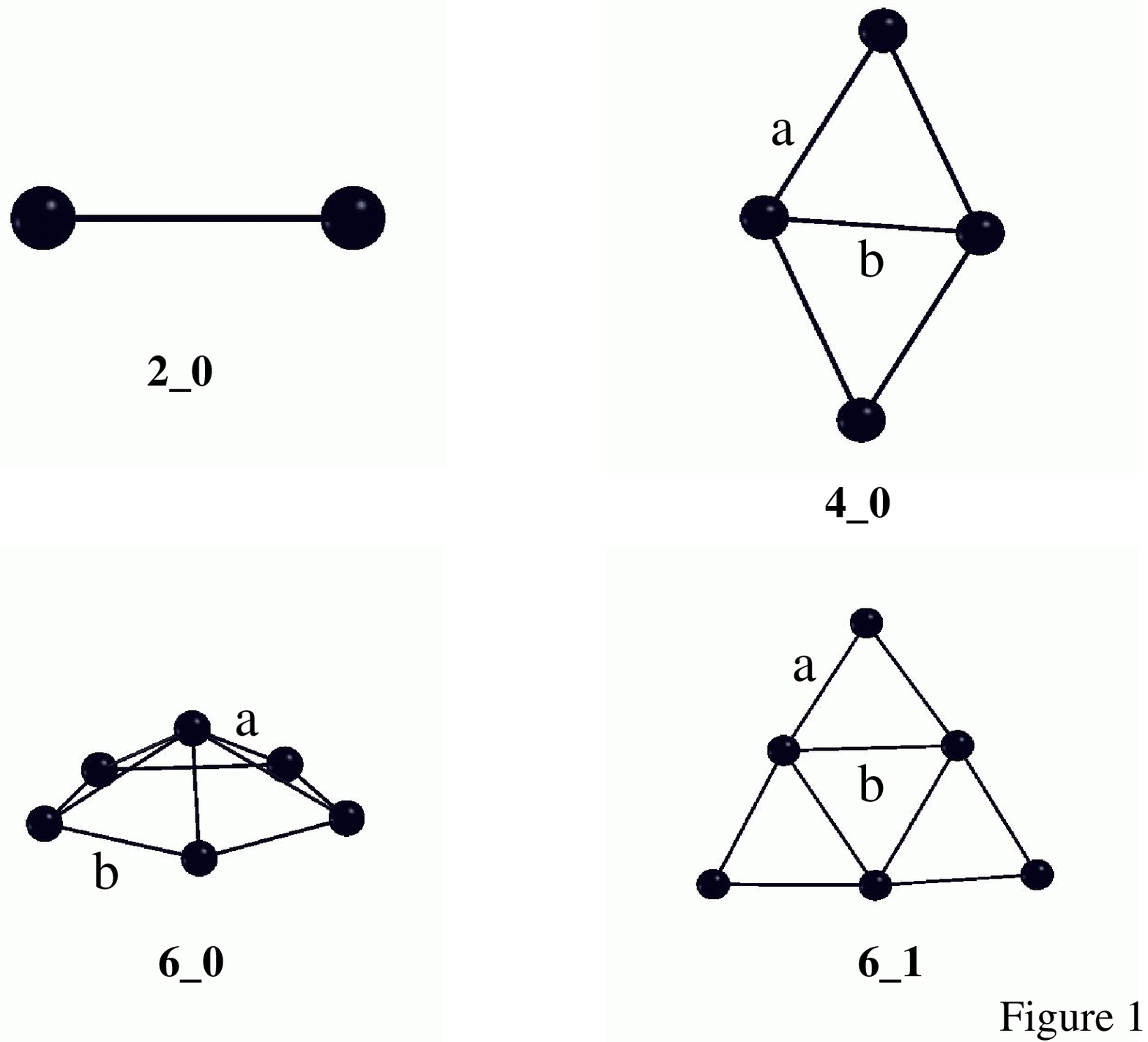}
\label{fig1}
\end{center}
\end{figure}
\begin{figure}
\begin{center}
\includegraphics{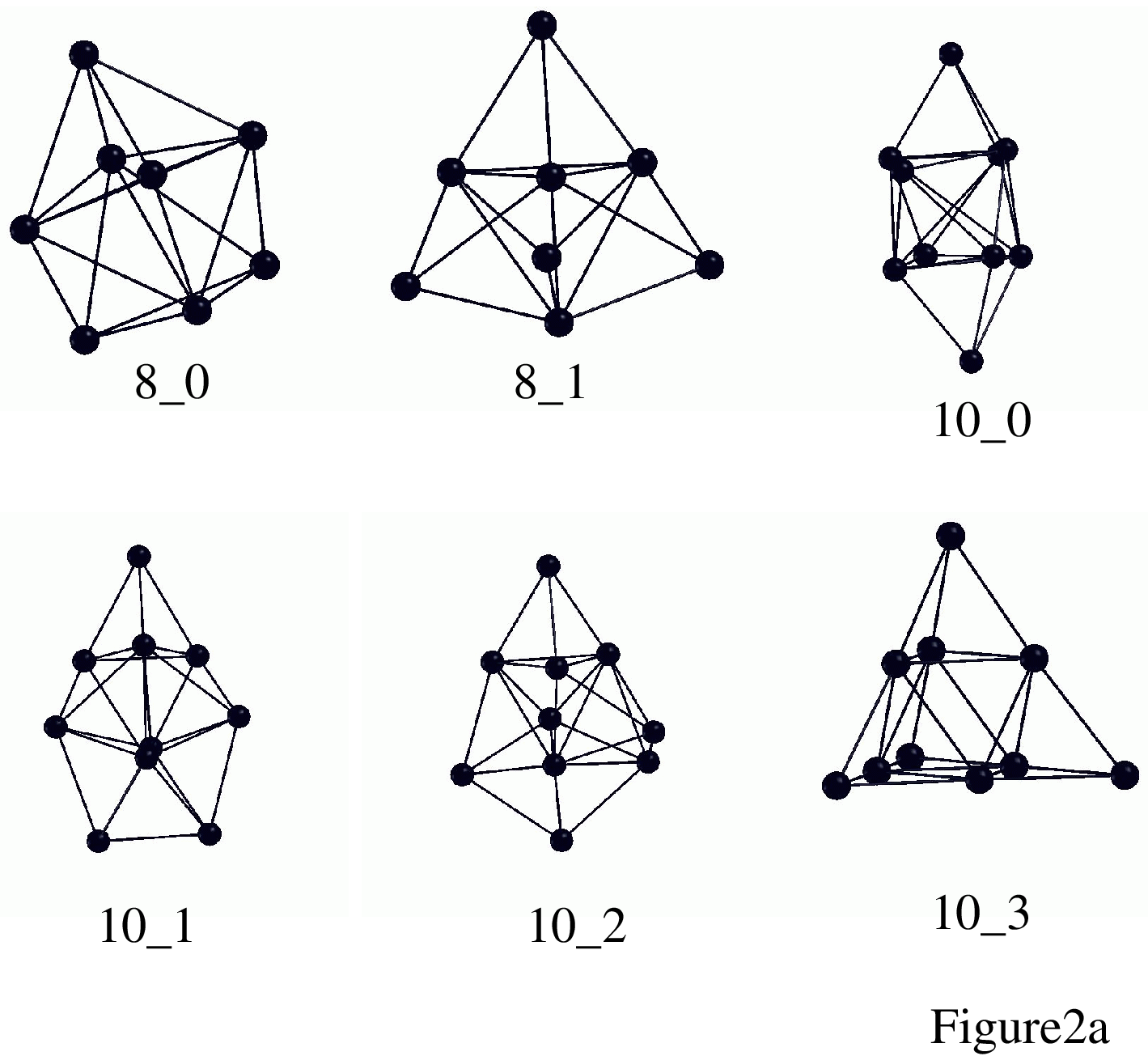}
\label{fig1}
\end{center}
\end{figure}
\begin{figure}
\begin{center}
\includegraphics{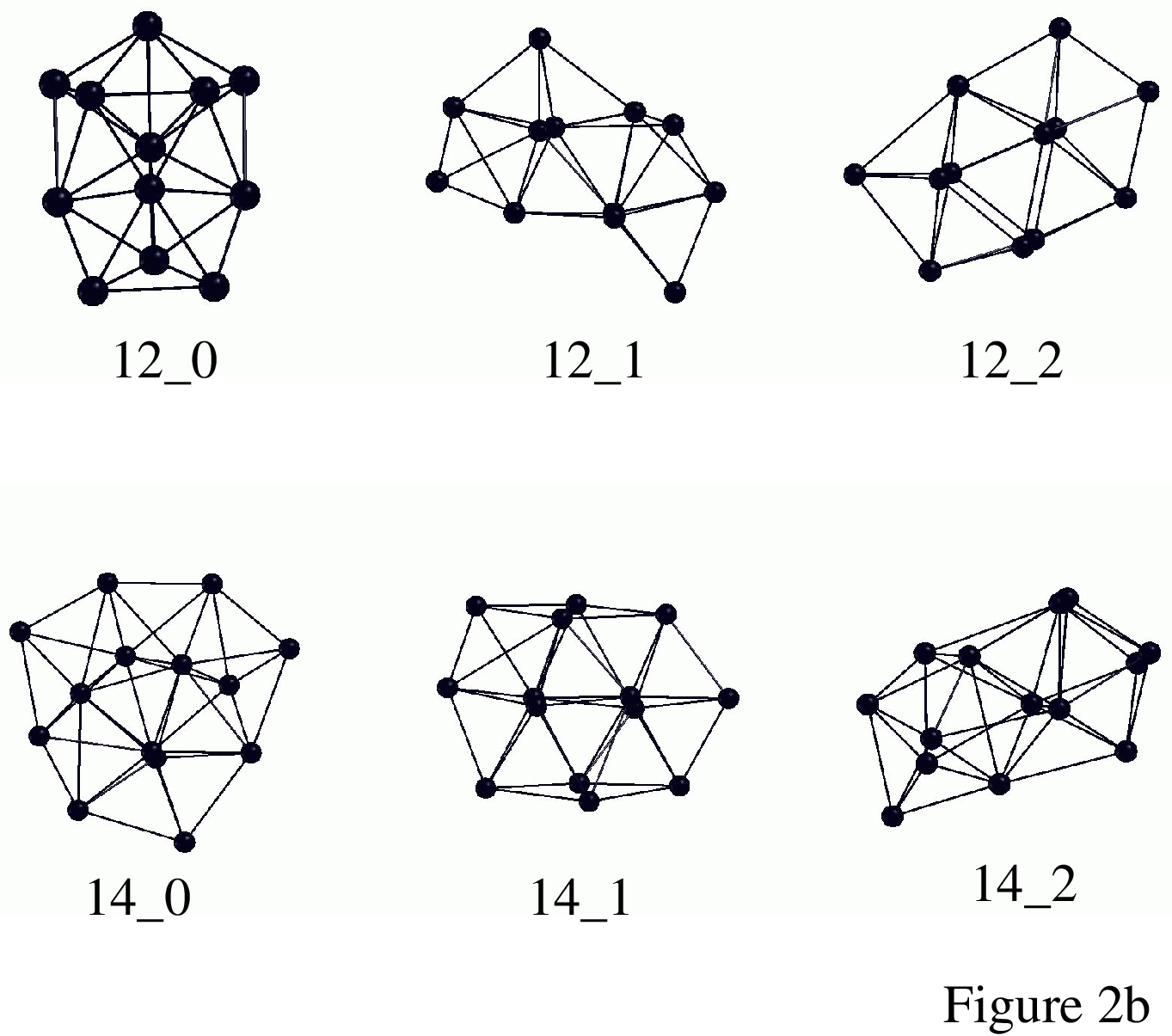}
\label{fig1}
\end{center}
\end{figure}
\begin{figure}
\begin{center}
\includegraphics{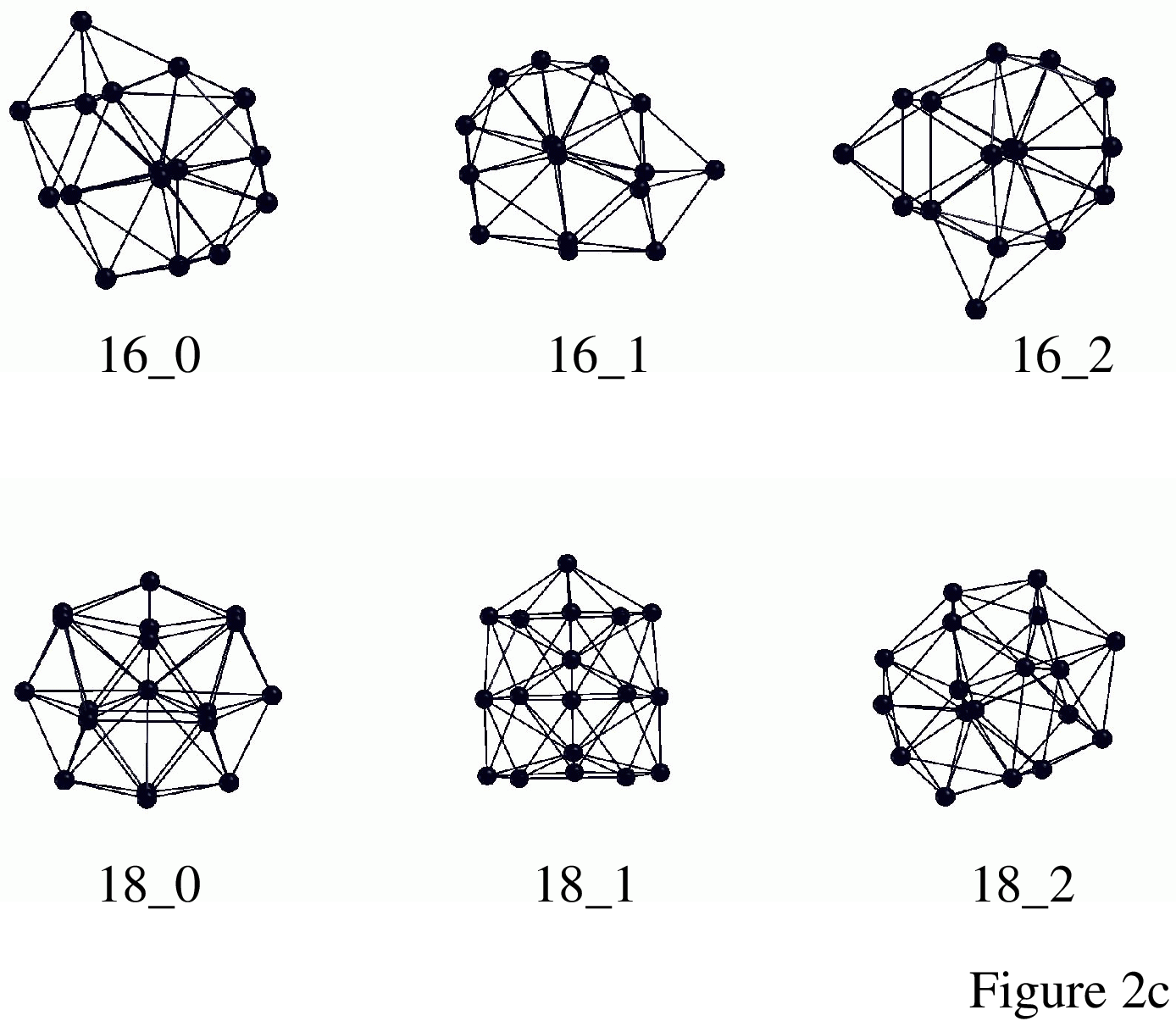}
\label{fig1}
\end{center}
\end{figure}
\begin{figure}
\begin{center}
\includegraphics{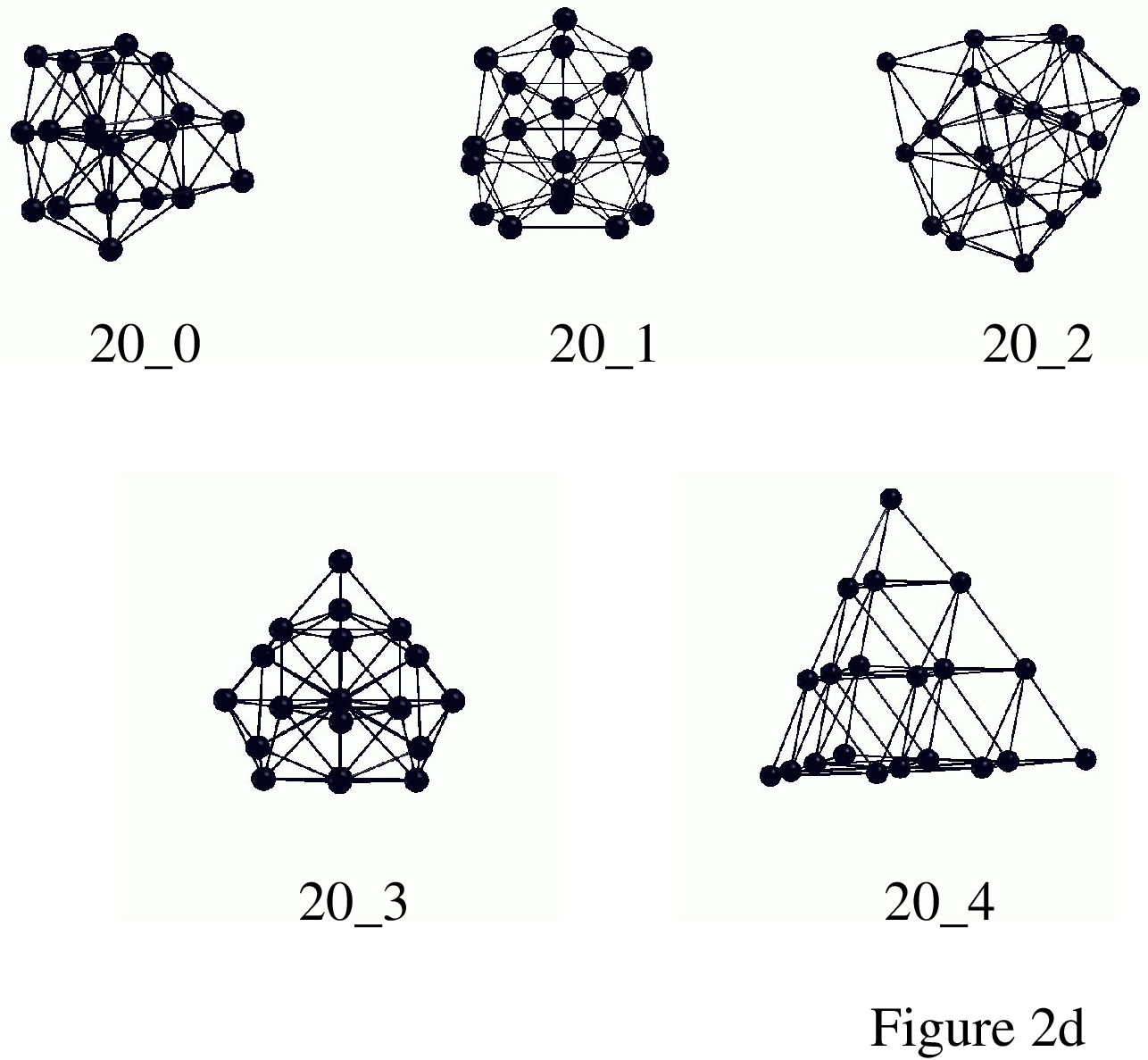}
\label{fig1}
\end{center}
\end{figure}
\begin{figure}
\begin{center}
\includegraphics{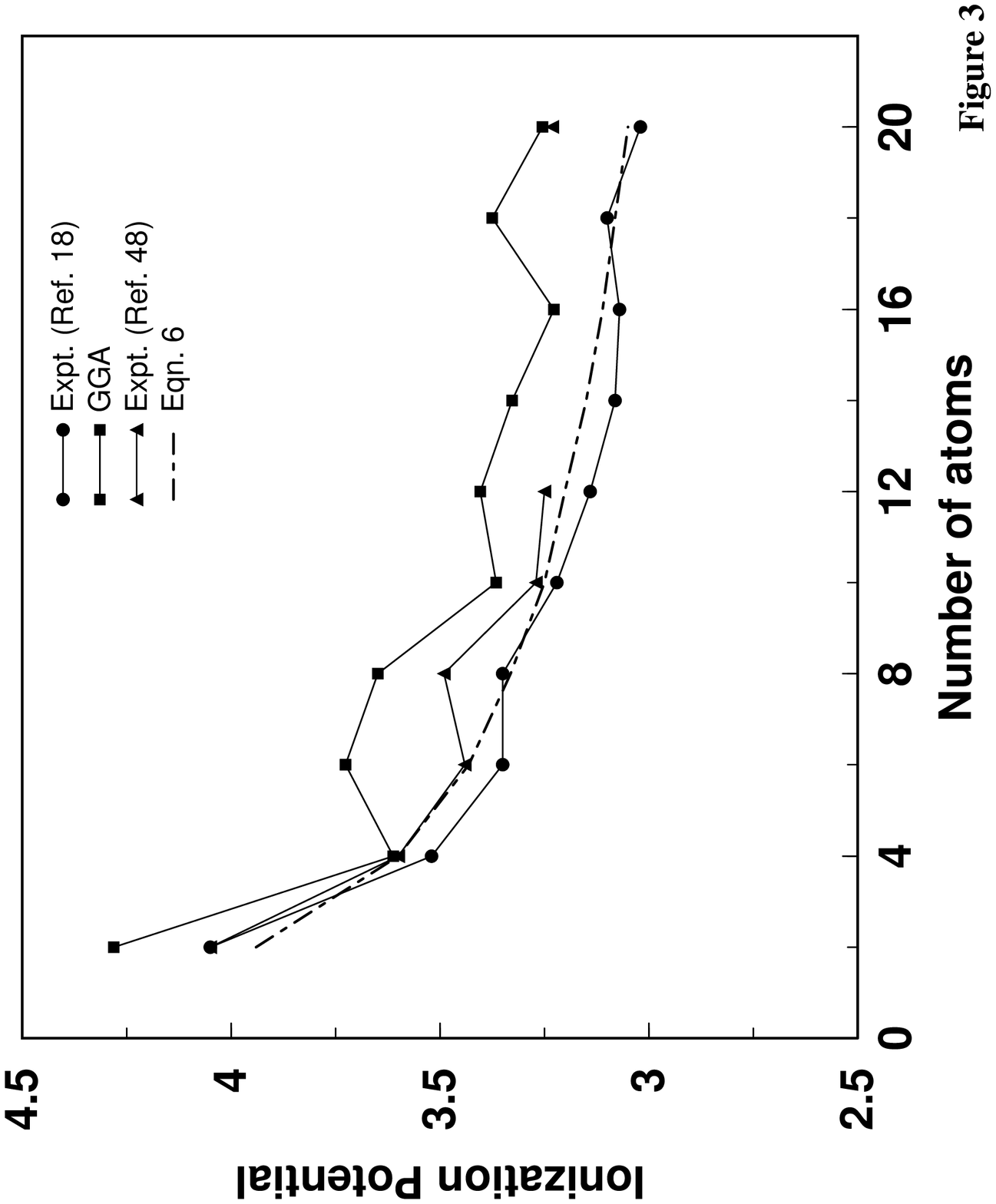}
\label{fig1}
\end{center}
\end{figure}

\begin{figure}
\begin{center}
\includegraphics{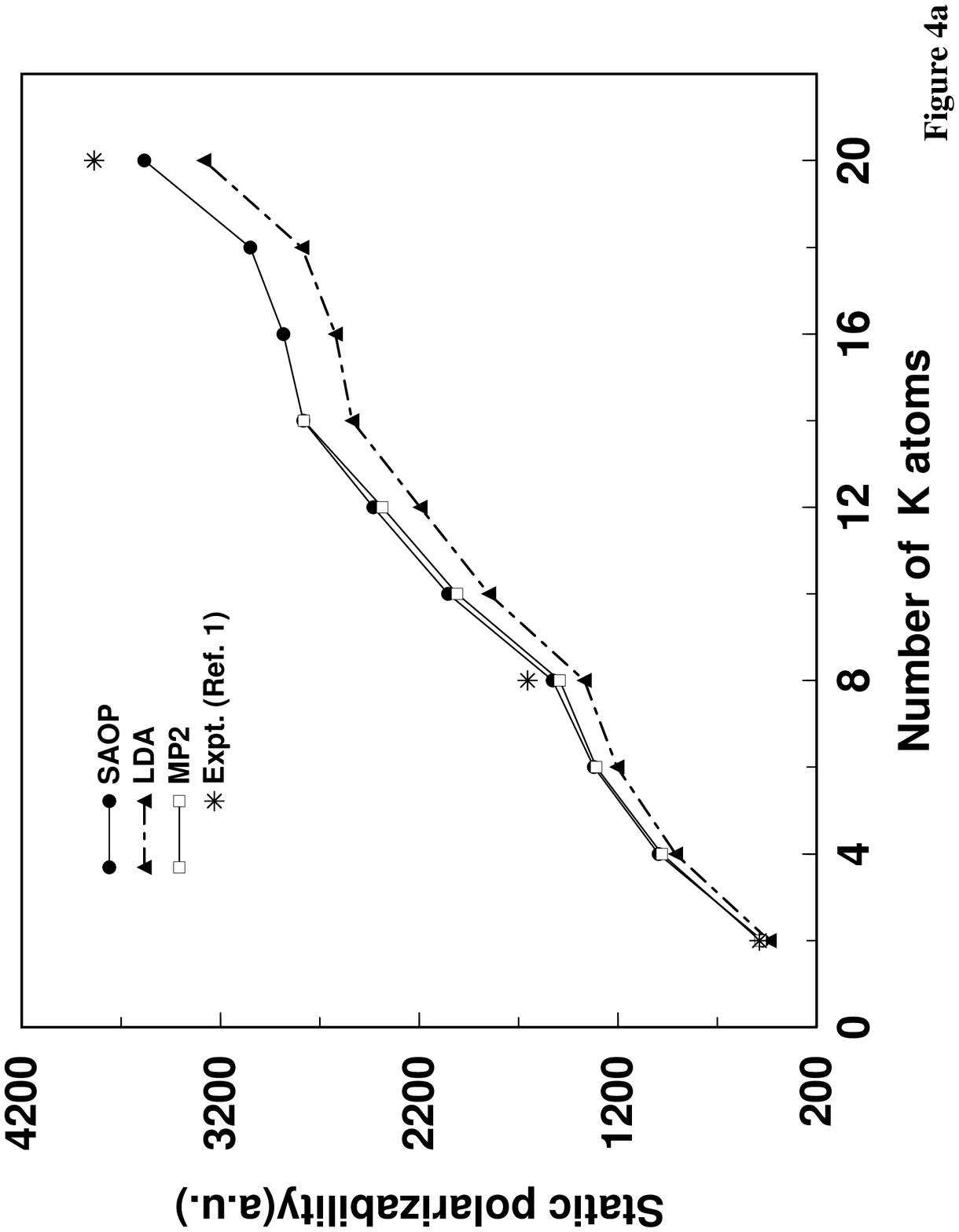}
\label{fig1}
\end{center}
\end{figure}

\begin{figure}
\begin{center}
\includegraphics{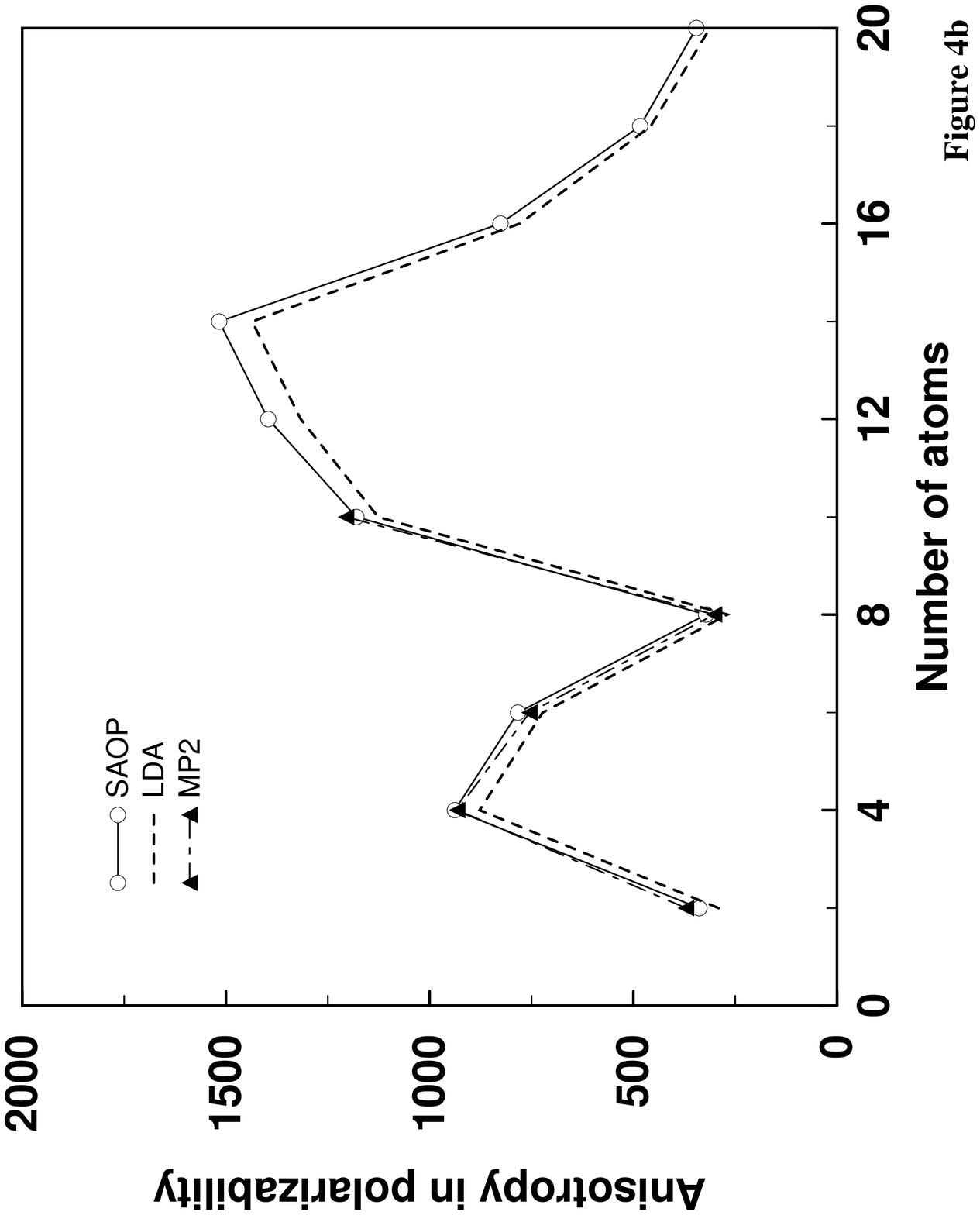}
\label{fig1}
\end{center}
\end{figure}

\begin{figure}
\begin{center}
\includegraphics{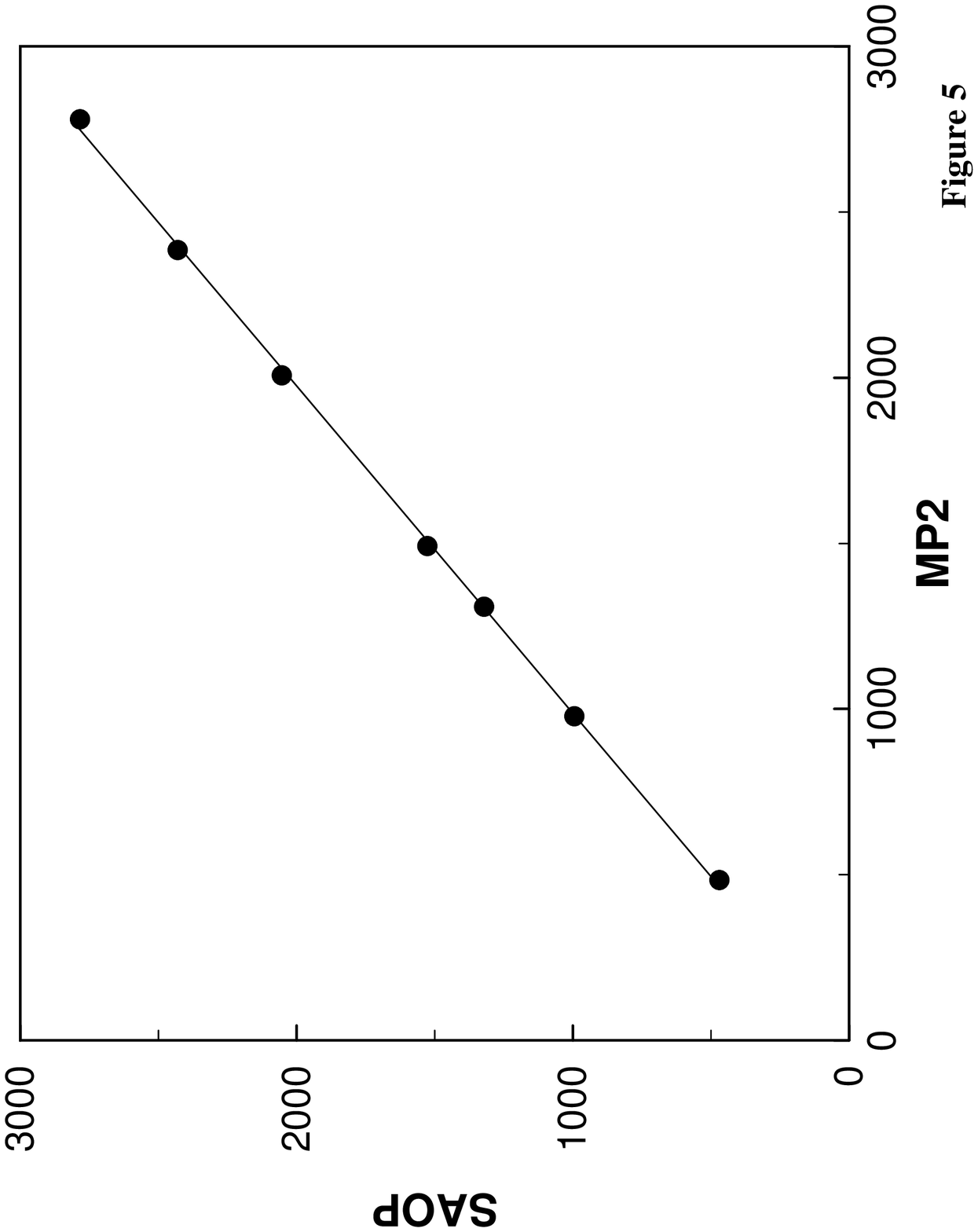}
\label{fig1}
\end{center}
\end{figure}

\begin{figure}
\begin{center}
\includegraphics{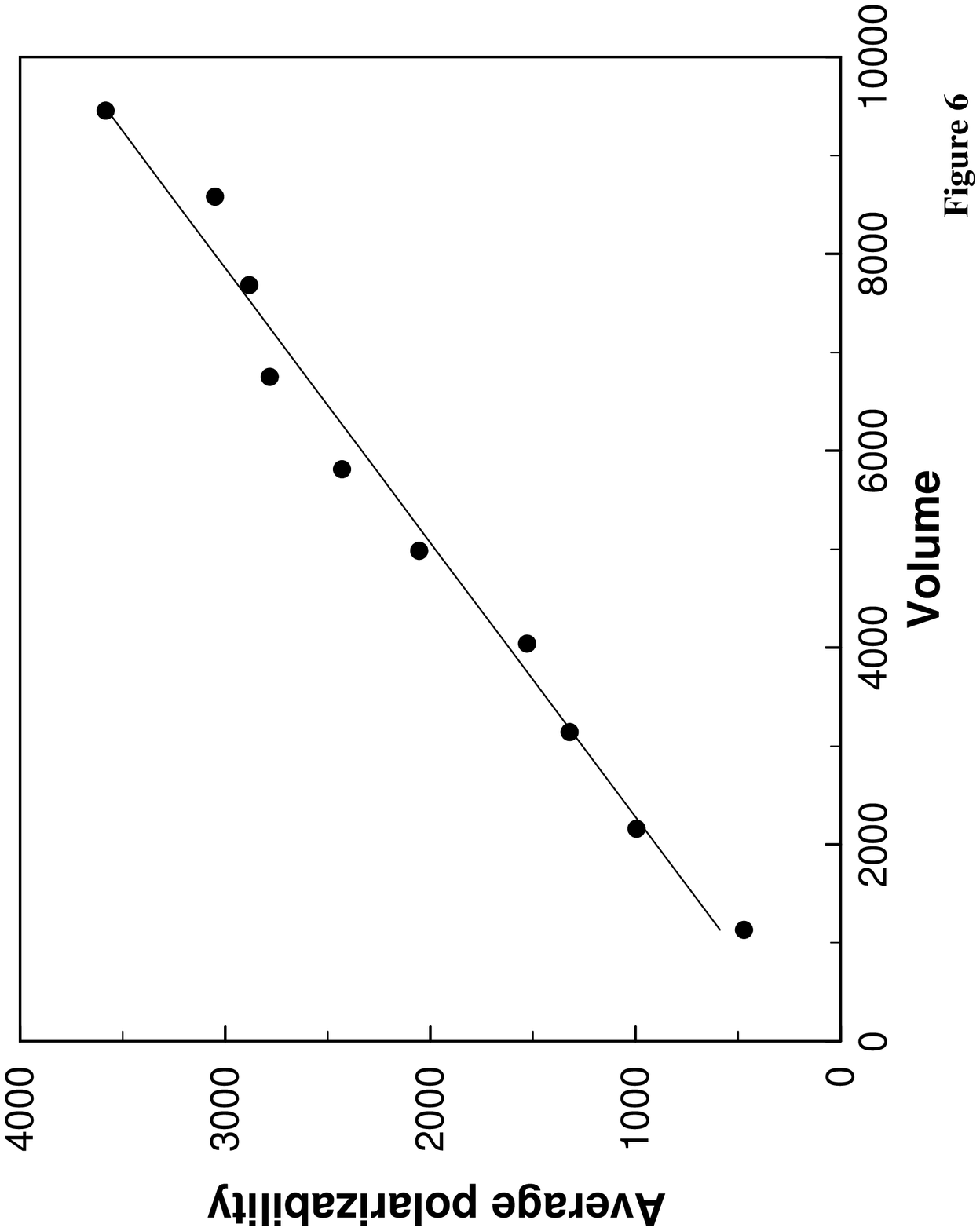}
\label{fig1}
\end{center}
\end{figure}
\end{document}